\documentstyle[aps,prl]{revtex}
\begin{document}

\twocolumn[\hsize\textwidth\columnwidth\hsize\csname
@twocolumnfalse\endcsname

\title{The origin and properties of the wetting layer and
early evolution of epitaxially strained thin films}
\author{Helen R. Eisenberg\cite{email1} and Daniel Kandel\cite{email2}}
\address{Department of Physics of Complex Systems,\\
Weizmann Institute of Science, Rehovot 76100, Israel}

\maketitle

\begin{abstract}
\newline
We showed that a wetting layer in epitaxially strained thin films which
decreases with increasing lattice mismatch strain arises due to the
variation of nonlinear elastic free energy with film thickness. We
calculated how and at what thickness a flat film becomes unstable to
perturbations of varying size for films with both isotropic and
anisotropic surface tension. We showed that anisotropic surface tension
gives rise to a metastable enlarged wetting layer. The perturbation
amplitude needed to destabilize this wetting layer decreases with
increasing lattice mismatch. We also studied the early evolution of
epitaxially strained films. We found that film growth is dependent on
the mode of material deposition. The growth of a perturbation in a flat
film is found to obey robust scaling relations. These scaling relations
differ for isotropic and anisotropic surface tension.
\newline
\end{abstract}

PACS numbers: 68.55.Jk, 81.15.Aa
\newline

]

\input epsf

\section{Introduction}

The growth of epitaxially strained thin films in which there is lattice
mismatch between the substrate and the film is of major importance in
the fabrication of semiconductor and optoelectronic devices. The
lattice mismatch generates strain in the deposited film, which can
cause film instability unfavorable to uniform flat film growth. The
strained film can relax either by the introduction of dislocations or
by the formation of coherent (dislocation-free) islands on the film
surface via surface diffusion. These coherent islands can self-organize
to create periodic arrays which can be utilized to create quantum dot
structures of electronic significance. Understanding and predicting
strained thin-film evolution is important for the improved fabrication
of semiconductor devices.

Early film growth tends to occur via coherent island formation as there
is an energy barrier to the introduction of dislocations. Dislocations
occur at island edges once islands reach a certain size as the large
stress at island edges provides a pathway for dislocation formation
\cite{eaglesham}. We only consider dislocation-free films since many
experiments show an absence of dislocations (see e.g
\cite{eaglesham,mo}) especially in early film evolution.

The current work sheds light on the following two problems: First, it
has been observed experimentally that dislocation-free flat films of
less than a certain thickness (the critical wetting layer) are stable
to surface perturbations, while thicker films are unstable
\cite{mo,massies,ramachandran,kamins,floro1997,floro1999,tromp,perovic,osten}.
The thickness of the wetting layer is substance dependent and decreases
with increasing lattice mismatch strain
\cite{floro1997,floro1999,tromp,perovic,osten},
$\varepsilon=(a_{s}-a_{f})/a_{f}$, where $a_{s}$ and $a_{f}$ are the
substrate and film lattice constants. Above the critical wetting layer,
3D coherent islands form. Despite considerable efforts the physics of
the critical wetting layer is poorly understood. Namely, why is there a
critical, stable wetting layer and what controls its thickness? As in
most cases heteroepitaxial growth is done below the roughening
transition, how does anisotropic surface tension affect the thickness
of the critical wetting layer? The second question we address is how
does continuous material deposition affect the early evolution of thin
films?

 Previous works about the existence and nature of the critical wetting layer
 can be split into two
categories: those which looked at the dynamic stability of a flat film
to small perturbations \cite{chiu,spencer}, and those which looked at
whether a flat film is energetically favourable to a film with fully
formed faceted islands \cite{tersoff,daruka,wang,roland}.

The research in the first category addressed substances with isotropic
surface tension. In these works physical parameters (lattice mismatch
or surface tension) which differ between the substrate and film were
smoothly varied over the substrate-film interface in order to avoid
nonanalyticities. The effect of such smoothing was to create a wetting
layer. Whilst the choice of a smoothing length of the order of a
lattice parameter is physically reasonable, none of these works gave a
physical explanation for the smoothing of material parameters over the
interface or tried to physically calculate the smoothing length or form
of the transition.

The research in the second category typically used physically motivated
methods in order to determine the free energy of a flat film of varying
depth. Tersoff \cite{tersoff} and Roland and Gilmer \cite{roland} both
used empirical potential methods to determine the chemical energy of
flat films of Ge/Si(001). Tersoff \cite{tersoff} then compared this
chemical potential with that of a bulk strained Ge in order to
determine whether it is preferable to form islands and to predict a
wetting layer of 3 monolayers. Roland and Gilmer \cite{roland} saw some
evidence of clustering in thicker films in molecular dynamics
simulations. Daruka and Barabasi \cite{daruka} used an expression for
the free energy of a flat film of varying depths which fits the results
of the earlier works \cite{tersoff,roland} in order to compare the
energies of flat films and films with fully faceted islands whose
energies were calculated using continuum elasticity. They saw a wetting
layer which increased with decreasing lattice mismatch. Wang et al.
\cite{wang} used ab initio methods in order to determine the formation
energy of flat films of varying depths of InAs/GaAs(100). They compared
this energy with that of a thinner film with fully faceted islands
whose energies were calculated using continuum elasticity. All the
above works did not study the issue of when a flat film becomes
unstable to small monolayer perturbations or the dynamics of growth.

In this paper we show that the variation of nonlinear elastic free
energy with film thickness can give rise to a wetting layer which
decreases with increasing lattice mismatch strain. We show how and at
what depth a flat film becomes unstable to perturbations of varying
size for films with both isotropic and anisotropic surface tension.
This provides a more realistic estimate of critical wetting layer
thickness than the studies described above, for films in which islands
grow from small surface perturbations rather than being immediately
nucleated on the flat film. This mode of growth has been seen in many
experimental systems \cite{floro1999,tromp,perovic,pidduck}, especially
for films with small lattice mismatch, $\varepsilon<2.5\%$. As
discussed below we study the evolution of these small perturbations and
observe island faceting.
 We show that anisotropic surface tension gives
  rise to a metastable enlarged
wetting layer. The perturbation amplitude needed to destabilize this
wetting layer decreases with increasing lattice mismatch.

The effects of material deposition on early thin film evolution was
addressed by Chiu and Gao \cite{chiu} who looked at the evolution of
strained films with isotropic surface tension when material deposition
is constant in the direction perpendicular to the film surface,
corresponding to liquid phase epitaxy. In the present paper we look at
thin film growth when material deposition is constant in the direction
perpendicular to the film surface and when deposition is at a steady
rate in the $y$ direction (vertical to the interface between the film
and the substrate), corresponding to any directed deposition (e.g,
molecular beam epitaxy). The latter is a much more common method of
material deposition in strained film growth. We found that the type of
evolution seen depends on the direction of material deposition. When
the deposition is constant in the vertical $y$-direction, the film
evolves according to the linear evolution equation, even after the
surface is no longer a sine function and cusp formation occurs. When
deposition is constant perpendicular to the surface,
 cusp formation is slowed down at
very high deposition rates and the surface shows signs of reaching a
steady-state morphology. We also studied thin film evolution for
faceting films, and found
 robust scaling laws for film growth.

The rest of the paper is organized as follows. In Sec. II we formulate
the problem. In Sec. III we present the general results of linear
stability analysis. In Sec. IV we describe how we calculated the
variation of the nonlinear elastic free energy of a flat film with film
thickness. It is this variation which gives rise to the wetting layer.
Calculations were carried out using a ball and spring model in order to
determine general qualitative behaviour. Sec. V describes the results
of the numerical simulations of thin film evolution without material
deposition, and in particular the metastability of the wetting layer.
Sec. VI describes the results of the numerical simulations of thin film
growth with material deposition. Some of the results on thin film
growth without deposition and a brief description of the
ball-and-spring model for calculating the free energy of a flat film
appear in our earlier paper \cite{eisenberg}.

\section{Problem Formulation}

We model the evolution of a thin film on a substrate using continuum
theory. The lattice mismatch between the film and the substrate creates
a strain in the film, $\varepsilon$ . Both the substrate and the film
are assumed to be elastically isotropic with the same elastic
constants. The surface of the solid is at $y=h(x,t)$ and the film is in
the $y>0$ region with the film-substrate interface at $y=0$. The system
is modelled to be invariant in the $z$-direction, and all quantities
are calculated for a section of unit width in that direction. This is
consistent with plane strain where the solid extends infinitely in the
$z$ direction and hence all strains in this direction vanish, i.e.
$e_{xz}=e_{yz}=e_{zz}=0$. We assume there is no material mixing between
the substrate and the film.

\begin{figure}[h]
   \epsfxsize=80mm
   \centerline{\epsffile{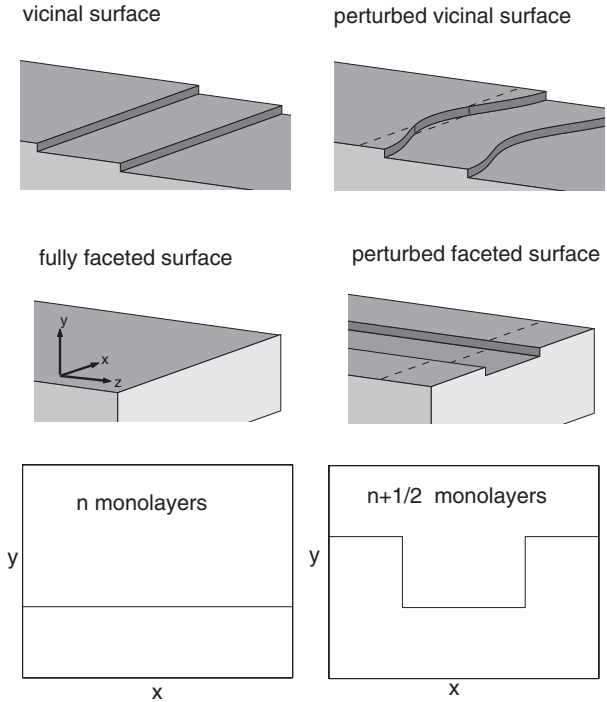}}
   \vspace{0.5cm}
\caption{Perturbations of vicinal and fully faceted surfaces. The
dotted lines represent the cross-sections taken in the $xy$ plane which
are shown in the bottom graphs. The figure shows that nucleation of new
steps is needed in order to perturb a facet, but not in order to
perturb a surface vicinal in the $z$ direction. }
    \label{vicinal}
\end{figure}

All the results mentioned in this paper relate to vicinal surfaces with
a very small miscut angle in the $z$ direction (see Fig.
\ref{vicinal}). Experimentally, surfaces often have such a small
miscut, as it is very difficult to grow a perfect facet. In such
surfaces there is no finite energy barrier for the formation of an
infinitesimal perturbation on the surface, since such a perturbation
involves only step motion and bending, and no nucleation of new steps.
For a faceted surface with no miscut in the $z$ direction there is a
finite energy barrier for the formation of an infinitesimal
perturbation associated with nucleation of step pairs.

The continuum approximation in the lateral direction ($x-z$ plane) is
valid as the film is infinite in the $z$ direction and the smallest
lateral surface features we study have widths of tens of atoms. However
in the vertical $y$ direction the films we are studying are sometimes
only a few monolayers thick. Is the continuum model valid for such a
film? Can inherently discrete system properties, such as the change in
free energy as a monolayer is added, be interpolated to films of a
non-integer number of monolayers?

 First, consider fully faceted films. As can
be seen in Fig. \ref{vicinal} there is a qualitative difference between
films with integer and non-integer numbers of monolayers, and so
discrete system properties cannot be interpolated to films with a
non-integer number of monolayers. Hence for such films continuum models
are not expected to give accurate results. To accurately model fully
faceted film growth, the discrete nature of the steps needs to be taken
into account using atomistic models and simulations. Modeling large
systems in such a manner is currently beyond computational
capabilities. However, as can be seen in Fig. \ref{vicinal}, the above
arguments do not apply to vicinal surfaces. In this case a perturbation
changes the morphology in a smooth manner without any nucleation of
steps. Therefore it is possible to estimate the properties of films
with a non-integer number of monolayers by interprolating the results
of films of integer numbers of monolayers. Thus, a continuum model
should adequately describe thin film growth behaviour for slightly
vicinal surfaces.

We assume that surface diffusion is the dominant mass transport
mechanism. Gradients in the chemical potential produce a drift of
surface atoms with an average velocity, $v$, given by the
Nernst-Einstein relation

\begin{equation}
v=-\frac{D_{s}}{k_{B}T}\frac{\partial \mu }{\partial s}, \label{nernst}
\end{equation}
where $D_{s}$ is the surface diffusion coefficient, $s$ is the arc
length, $T$ is the temperature, $k_{B}$ is Boltzmann's constant and
$\mu $ is the chemical potential at the surface; i.e. it is the
increase in free energy when an atom is added to the solid surface at
the point of interest. Taking the divergence of the surface current
produced by the atom drift gives an expression for the surface
movement\cite{mullins}

\begin{equation}
\frac{\partial h}{\partial t}=\frac{D_{s}\eta \Omega}{k_{B}T}
\frac{\partial }{\partial x}\frac{\partial \mu }{\partial s},
\label{evol}
\end{equation}
where $\eta $ is the number of atoms per unit area on the solid surface
and $\Omega$ is the atomic volume.

In the continuum approximation $\mu =\Omega \frac{\delta F}{\delta h}$,
where $F$ is the free energy of the solid and $\delta F/\delta h$ is
the functional derivative of $F$. The free energy is composed of
elastic and surface terms:

\begin{equation}
F=F_{el}+\int dx\ \gamma \sqrt{1+(\partial h/\partial x)^{2}},
\label{free energy}
\end{equation}
where $\gamma $ is the surface tension and $F_{el}$ is the elastic free
energy including any elastic contributions to the surface tension.

 In general the elastic free energy can be written as
 $F_{el}=F_{el}^{(0)}+\delta F_{el}$, where $F_{el}^{(0)}$ is the
 elastic free energy of a zero strain reference state.
  The elastic free energy can be written in terms
 of the elastic free energy density, $f_{v}$, as
 $F_{el}=\int dxdyf_{v}$.
$f_{v}$ is expanded as a power series in the strain: $f_{v}=f_{v}^{(0)}
+\sigma _{ij}^{(0)}e_{ij}+\frac{1}{2} c_{ijkl}e_{ij}e_{kl}+\cdots $,
where $f_{v}^{(0)}$ is the free energy density in the zero strain
reference state, $\sigma _{ij}^{(0)}$ is the stress in the reference
state and $c_{ijkl}$ are the elastic coefficients of the material. In
linear elasticity theory, deformations are assumed to be small and so
terms of third order and higher are neglected. The stress-strain
relationship is given by $\sigma _{ij}=\frac{\partial f_{v}}{\partial
e_{ij}} $, which under linear elasticity gives Hooke's law:
\begin{equation}
 \sigma _{ij}=\sigma _{ij}^{(0)}+c_{ijkl}e_{kl}.
\label{stressstrain}
\end{equation}
 We now briefly describe the equations which
need to be satisfied in order to completely determine the equilibrium
stress and strain in an elastic body. An elastic solid must satisfy the
equations of mechanical equilibrium at every point in its interior (see
e.g. \cite{sokolnikoff}):

\begin{equation}
\partial _{j}\sigma _{ij}+\xi _{i}=0, \label{int equil}
\end{equation}
where $\xi _{i}$ is an external force on the solid. The solid must also
satisfy the equations of equilibrium at its surface,

\begin{equation}
\sigma _{ij}n_{j}=T_{i}^{n}, \label{surf equil}
\end{equation}
where $n$ is the exterior normal, and $T^{n}$ is the external force
acting on the unit area surface element with normal $n$.  As the
strains $e_{ij\text{ }}$ are not independent but are linked via the
displacements of the elastic body, they must also satisfy the equations
of compatibility:

\begin{equation}
\partial _{k}\partial _{l}e_{ij\text{ }}+\partial _{i}\partial _{j}e_{kl
\text{ }}=\partial _{j}\partial _{l}e_{ik\text{ }}+\partial
_{i}\partial _{k}e_{jl\text{ }}. \label{cont}
\end{equation}

Eqs. (\ref{int equil}),(\ref{surf equil}) and (\ref{cont}) along with
the stress-strain relationships (\ref{stressstrain}) give us a system
of equations, which are sufficient for the complete determination of
the equilibrium stress and strain in an elastic body.

For the system we study, external body forces (e.g. gravity) are
neglected. Hence Eq. (\ref{int equil}) becomes

\begin{equation}
\partial _{j}\sigma _{ij}=0\qquad \text{for} \; y<h(x). \label{equil}
\end{equation}
Our system has periodic boundary conditions in the $x$ direction and is
infinite in the negative $y$ direction. We shall assume that the forces
on the upper surface due to surface tension (as given by Marchenko and
Parshin\cite {Marchenko}) are negligible in comparison to the forces
due to the mismatch stress. This assumption is fulfilled as long as
$\frac{\gamma } {R}\ll M\varepsilon $ where $R$ is the radius of
curvature of the surface and $M$ is the\ plane strain modulus. For
typical values of $ \gamma$, $M$ and $\varepsilon $, this condition is
satisfied when $R$ is larger than the lattice constant. As typical
surface features have length scales of the order of $100nm$ this
assumption is valid. Hence the boundary conditions are given by

\begin{eqnarray}
\sigma _{ij}n_{j} &=0& \;\;\;\text{ at }\;\;\;\;\;\;\;\;   y=h(x)
\nonumber \\
 \sigma _{ij} &\rightarrow 0   & \;\;\;\text{ when
}\;\;\; y\rightarrow -\infty \label{bound}
\end{eqnarray}

We now return to our discussion on the determination of the elastic
free energy of both the reference state and the perturbed state. For
each value of $x$, our reference state corresponds locally to a
\textit{flat} film of thickness $h(x)$ constrained to have the lateral
lattice constant of the substrate; i.e., $F_{el}^{(0)}=\int
dx\int_{-\infty }^{h(x)}dyf_{v}^{(0)}(h(x),y)$, where
$f_{v}^{(0)}(h(x),y)$ is the elastic free energy density of a flat film
of thickness $h(x)$ with the substrate lateral lattice constant. We
calculate the correction to the elastic free energy of the perturbed
state, $\delta F_{el}$, using linear elasticity theory.

When looking at the stability of a strained flat film of thickness $C$,
the obvious first choice for a reference state is that of a flat film
of thickness $C$ constrained to have the lateral lattice constant of
the substrate. For later calculations we must fully define the
reference state and hence need to know its stress
$\sigma_{ij}^{(0)}(C,y)$ and free energy density $f_{v}^{(0)}(C,y)$.
One simple approach to calculate these quantities would be to use
linear elasticity with the unstressed film as a reference state. In
linear elasticity a flat film of any thickness constrained to have the
substrate lateral lattice constant and free to move in the $y$
direction is in equilibrium and has the elastic free energy density of
an infinitely strained film. Hence such a calculation does not predict
any $C$ or $y$ dependence in $\sigma_{ij}^{(0)}$ and $f_{v}^{(0)}$
except for a step function at the film-substrate interface. For example
in the case of plane strain where the mismatch strain is uniaxial
(i.e., $e_{xz}=e_{yz}=e_{zz}=0,e_{xy}=0,e_{xx}=\varepsilon$), linear
elasticity gives $\sigma_{ij}^{(0)}=M\varepsilon$ and
$f_{v}^{(0)}=M\varepsilon^{2}/2$, where $M$ is the plane strain
modulus. Therefore, variation of the elastic free energy and stress of
a flat film with film height is a nonlinear phenomenon, and a model
outside of linear elasticity theory must be used to calculate them. As
will be shown in Section. III, small variations in the reference free
energy density with film thickness are crucial in predicting wetting
layer thickness, and a reference free energy density which has no
variation with film thickness will lead to thin films that have no
wetting layer.

The disadvantage of our choice of the reference state is that the
dependence of $h$ on $x$ leads to lateral variations of the reference
state. As a result, the reference stress does not satisfy the condition
of mechanical equilibrium. However, the needed corrections vanish in
the limit $a/\lambda \rightarrow 0$, where $a$ is the length scale over
which stress varies in the $y$-direction and $\lambda $ is the lateral
length of typical surface structures. This is because in this limit
there are no lateral variations in the reference stress. As typical
experimental islands have $\lambda \sim 100nm$, and as $a$ is of the
order of the lattice constant (see below), the corrections to the
reference stress are small and have been ignored.

  Though linear elasticity
cannot be used to calculate properties associated with the reference
state, it can still be used to find the correction to the elastic free
energy of the perturbed state, $\delta F_{el}$. For convenience we work
in terms of the reference elastic free energy per unit length in the
$x$-direction, $f_{el}^{(0)}(h(x))\equiv \int_{-\infty
}^{h(x)}dyf_{v}^{(0)}(h(x),y)$, instead of the free energy per unit
volume. According to linear elasticity theory, $\delta F_{el}=\int
dx\int_{-\infty }^{h(x)}dy\left( \sigma
_{ij}^{(0)}e_{ij}+\frac{1}{2}c_{ijkl}e_{ij}e_{kl} \right) $. In terms
of the stress tensor, we find

\begin{eqnarray}
F_{el}=&\int& dx\ f_{el}^{(0)}+  \nonumber \\
&\int& dx\int_{-\infty }^{h(x)}dy\left( \frac{1}{2}S_{ijkl}\sigma
_{ij}\sigma _{kl}-\frac{1}{2}S_{ijkl}\sigma _{ij}^{(0)}\sigma
_{kl}^{(0)}\right),   \label{elastic free energy}
\end{eqnarray}
where we have used the inverted Hooke's law $e_{ij}=S_{ijkl}(\sigma
_{kl}-\sigma _{kl}^{(0)})$. $S_{ijkl}$ are the compliance coefficients
of the material.

Using Eqs. (\ref{free energy}) and (\ref{elastic free energy}) we
arrive at an expression for $\delta F/\delta h$

\begin{eqnarray}
\frac{\delta F}{\delta h}&=&\widetilde{\gamma }(\theta)\kappa + \frac{\
df_{el}^{(0)}}{dh} \nonumber \\
&+&\left. \left( \frac{1}{2}S_{ijkl}\sigma _{ij}\sigma
_{kl}-\frac{1}{2} S_{ijkl}\sigma _{ij}^{(0)}\sigma _{kl}^{(0)}\right)
\right|
_{y=h(x)} \nonumber \\
&+&\int_{-\infty }^{h(x)}dy\frac{\partial }{\partial h}\left(
\frac{1}{2}S_{ijkl}\sigma _{ij}\sigma _{kl}-\frac{1}{2}S_{ijkl}\sigma
_{ij}^{(0)}\sigma _{kl}^{(0)}\right),  \label{dfdh2}
\end{eqnarray}
where $\kappa $ is surface curvature, $\theta $ is the angle between
the normal to the surface and the $y$ direction, and $\widetilde{\gamma
}(\theta )=\gamma (\theta )+\partial ^{2}\gamma /\partial \theta ^{2}$
is the surface stiffness. The integrand in Eq. (\ref{dfdh2}) is the
change in the linear elastic free energy density as the surface profile
changes infinitesimally. Following Sokolnikoff \cite{sokolnikoff} we
now show that this term is second order in $\delta h$ in the absence of
external surface or body forces and so can be neglected. When the
surface profile changes, the strain in the body changes from $e_{ij}$
to $e_{ij}+e_{ij}^{\prime }$, where $e_{ij}^{\prime }$ is of order
$\delta h$. The elastic free energy density can be written as $f_{v}+
\Delta f_{v}=\frac{1}{2} c_{ijkl}(e_{ij}+e_{ij}^{\prime
})(e_{kl}+e_{kl}^{\prime })$, and the change in elastic free energy
density is  $\Delta f_{v}=c_{ijkl}e_{ij} e_{kl}^{\prime }+
\frac{1}{2}c_{ijkl}e_{ij}^{\prime }e_{kl}^{\prime }$. Using Hooke's law
(\ref{stressstrain}), the definition of strain and Eq. (\ref{int
equil}), we rewrite $\Delta f_{v}$ as $\Delta f_{v}=\partial
_{j}(\sigma _{ij}u_{i}^{\prime }) +\xi _{i}u_{i}^{\prime
}+\frac{1}{2}c_{ijkl}e_{ij}^{\prime }e_{kl}^{\prime }.$ The total
change in the elastic free energy is
 $\int $ $\Delta f_{v}\ dxdy=\int
T_{i}^{n}u_{i}^{\prime }ds+\int \xi _{i}u_{i}^{\prime }dxdy+\int
\frac{1}{2}c_{ijkl}e_{ij}^{\prime }e_{kl}^{\prime }dxdy$, where the
first term on the r.h.s. is an integral over the film surface. To
obtain this we used Eq. (\ref{surf equil}). In the absence of external
surface or body forces the first two terms on the r.h.s of the above
equation vanish, and we are left with the equation $ \int $ $\Delta
f_{v}\ dxdy=\int \frac{1}{2}c_{ijkl}e_{ij}^{\prime }e_{kl}^{\prime
}dxdy$. $e_{ij}^{\prime }$ is of order $\delta h$, and hence the last
term in Eq. (\ref{dfdh2}) can be ignored for infinitesimal changes to
the solid surface and we have

\begin{eqnarray}
\frac{\delta F}{\delta h}&=&\widetilde{\gamma }\kappa +\frac{\
df_{el}^{(0)}}{dh} \nonumber \\
&\;&+\left. \left( \frac{1}{2}S_{ijkl}\sigma _{ij}\sigma
_{kl}-\frac{1}{2} S_{ijkl}\sigma _{ij}^{(0)}\sigma _{kl}^{(0)}\right)
\right| _{y=h(x)}. \label{dfdh3}
\end{eqnarray}

As the above equation gives $\frac{\delta F}{\delta h}$ at the solid
surface, all variables in the equation are also given at the surface.
In particular $\sigma _{ij}^{(0)}(h,y=h)$ is taken as the stress at the
surface of a flat solid of height $h(x)$ and hence must vanish when
$h\leq 0$, since then the film is absent. $df_{el}^{(0)}(h)/dh$ is
determined by calculating how the reference elastic free energy of the
solid changes as monolayers are added to the solid surface. When $h\leq
0,$ $ df_{el}^{(0)}/dh=0$ as the substrate is completely relaxed. In
principle, Eq. (\ref{dfdh3}) should also contain derivatives of $\gamma
$ with respect to $h$. However, we believe that the variation of
surface tension with $h$ away from a step dependence is due to elastic
effects. Since we included all elastic contributions in the zero-strain
elastic free energy, we modeled $\gamma $ as a step function, taking
the value of the substrate surface tension for $h\leq 0$ and the film
surface tension for $h>0$. Thus all partial derivatives of $\gamma $
with respect to surface height vanish and were omitted from Eq.
(\ref{dfdh3}).

Equations (\ref{evol}) and (\ref{dfdh3}) form a complete model of
surface evolution. In order to solve this model, the chemical potential
(given by Eq. (\ref{dfdh3})) for a given surface must be found, and so
the linear elastic free energy density at the solid surface,
$f_{v}^{lin}=\left. \frac{1}{2} S_{ijkl}\sigma _{ij}\sigma _{kl}\right|
_{y=h(x)}$ must be calculated. For an isotropic solid under plane
strain with zero force boundary conditions the above
 expression simplifies considerably to give, $f_{v}^{lin}=
\left.\frac{1}{2M}(\sigma _{xx}+\sigma _{yy})^{2}\right| _{y=h(x)}$,
where $M$ is the plane strain modulus. Hence we must determine the
stress at the surface of the film. This is done by finding the stress
which satisfies both the linear elasticity equations (the equations of
compatibility(\ref{cont}) and equilibrium(\ref{equil})) and the
boundary conditions (\ref{bound}).

For an isotropic solid under plane strain, finding the stress which
satisfies the linear elasticity equations and the boundary conditions
can be reduced to finding the stress function, $W,$ which satisfies the
boundary conditions (\ref{bound}) and the biharmonic equation (see e.g.
Timoshenko\cite{timoshenko} or Mikhlin\cite {Mikhlin}):

\begin{equation}
\Delta ^{2}W=\Delta (\Delta W)=\frac{\partial ^{4}W}{\partial
x^{4}}+2\frac{
\partial ^{4}W}{\partial x^{2}\partial y^{2}}+\frac{\partial ^{4}W}{\partial
y^{4}}=0, \label{biharmonic}
\end{equation}
with $\sigma _{xx}=\frac{\partial ^{2}W}{\partial y^{2}},\quad \sigma
_{xy}=-\frac{\partial ^{2}W}{\partial x\partial y}\quad$and$
\quad\sigma _{yy}=\frac{
\partial ^{2}W}{\partial x^{2}}.$

In order to model the early evolution of faceted islands, and to study
the effect of an anisotropic form of surface tension on the wetting
layer, we used the cusped form of surface tension given by Bonzel and
Preuss \cite{bonzel}, which shows faceting in a free crystal: $\gamma
(\theta )=\gamma _{0}\left[ 1+\beta \left| \sin (\pi \theta /(2\theta
_{0}))\right| \right] $, where $\beta \approx 0.05$ and
$%
\theta _{0}$ is the angle of maximum $\gamma $. The value of $\gamma
_{0}$ was taken as 1 J/m$^{2}$ in the substrate and about 75\% of that
in the film (as is the case for Si/Ge). This ensures a wetting layer of
at least one monolayer. We considered a crystal which facets at
0$%
^{\circ },\pm 45^{\circ }$ and $\pm 90^{\circ }$ with $\theta _{0}=\pi
/8$. The cusp gives rise to $\widetilde{\gamma }=\infty $. However, a
slight miscut of the low-index surface along the $z$ direction leads to
a rounding of the cusp, which can be described by
\begin{equation}
\gamma (\theta )=\gamma_0\left(1+\beta \sqrt{\sin^2 (\frac{\pi
}{2\theta _{0}}\theta )+G^{-2}}\right)~,  \label{surf ten}
\end{equation}
where, for example, $G=500$ corresponds to a miscut angle, $\Delta
\theta\approx 0.1^{\circ}$. As mentioned earlier all the results
mentioned in this paper relate to surfaces with a very small miscut
angle in the $z$ direction.

\section{Linear stability analysis}

In this section we carry out a linear stability analysis of Eq.
(\ref{evol}) against a sinusoidal perturbation of wavenumber $k$,
similar to that carried out in Ref. \cite{asaro} for an infinitely
thick stressed film. We thus look for a height profile of the form,
$h(x,t)=C+\delta(t)\sin kx$, which solves Eq. (\ref{evol}) to first
order in $\delta$. To calculate the linear elastic energy we find the
solutions of (\ref{biharmonic}), which satisfy the boundary conditions
(\ref{bound}). $\sigma_{xy}^{(0)}$ vanishes because the film is
hydrostatically strained, and $\sigma_{yy}^{(0)}=0$ since in the
reference state the force on the surface vanishes. Hence the only
non-zero component of the reference stress is $\sigma_{xx}^{(0)}(h,y)$.
Stress functions of the form
\begin{equation}
W=\sigma_{xx}^{(0)}(h,y)y^{2}/2+(A+By)e^{ky}sin(kx)
\end{equation}
satisfy the biharmonic equation. This gives stresses of the form
\begin{eqnarray*}
 \sigma_{xx}&=&k[2B+(A+By)k]e^{ky} \sin(kx)+
\sigma_{xx}^{(0)}(h,y)
 \\
 \sigma_{yy}&=&-k^{2}(A+By)e^{ky}sin(kx)  \\
\sigma_{xy}&=&-k[B+(A+By)k]e^{ky}\cos(kx).
\end{eqnarray*}

 To first order in $\delta$ the
stresses that satisfy the biharmonic equation and the boundary
conditions are given by:
\begin{eqnarray*}
 \sigma_{xx}&=&- \delta ke^{-kC}
\sigma_{xx}^{(0)}(C,C)[2+(y-C)k] e^{ky} \sin(kx) \\
&\;&+\sigma_{xx}^{(0)}(h,y)
 \\
 \sigma_{yy}&=&k^{2}e^{-kC}\sigma_{xx}^{(0)}(C,C)(y-C)e^{ky}sin(kx)  \\
\sigma_{xy}&=&\delta
ke^{-kC}\sigma_{xx}^{(0)}(C,C)[1+(y-C)k]e^{ky}\cos(kx).
\end{eqnarray*}
At the surface these stresses take the form:
 \begin{eqnarray*}
  \sigma_{xx}&=&-2\delta
k\sigma_{xx}^{(0)}(C,C)\sin(kx) \\
&\;&+\sigma_{xx}^{(0)}(C,C)+\delta\sin(kx)d\sigma_{xx}^{(0)}/dh|_{h=C}
 \\
 \sigma_{yy}&=&0  \\
\sigma_{xy}&=&\delta k\sigma_{xx}^{(0)}(C,C)\cos(kx).
\end{eqnarray*}
Note that all the derivatives in the Taylor expansions used in this
analysis are with respect to $h$, the reference film thickness and not
with respect to $y$, the depth within the reference film. This is
because in calculating the chemical potential we are interested in how
the free energy of the film changes as material is added or removed
from the film surface; i.e. how the free energy of the film changes as
the film thickness changes and not how the free energy density of the
film varies within the film.

 Using the above stresses in Eq. (\ref{dfdh3}), we obtain the
 expression
  \begin{eqnarray}
\frac{\delta F}{\delta h}&=&\widetilde{\gamma }\kappa +
\frac{\ df_{el}^{(0)}}{%
dh}+\frac{1}{2M}(\sigma _{xx}+\sigma _{yy})^{2}-\frac{1}{2M} (\sigma
_{xx}^{(0)}
+\sigma _{yy}^{(0)})^{2} \nonumber \\
&=&\delta\sin(kx)\left[\frac{\ d^{2}f_{el}^{(0)}}{%
dh^{2}}-2k\frac{(\sigma_{xx}^{(0)}(C,C))^{2}}{M}+\widetilde{\gamma
}_{0}k^{2}\right] \nonumber \\
&\;&+\frac{\ df_{el}^{(0)}}{dh}(C,C), \label{lin equil}
\end{eqnarray}
where $\widetilde{\gamma}_{0}=\widetilde{\gamma}(\theta=0)$. Combining
the above equation with the evolution equation (\ref{evol}) gives the
following equation for $\delta (t)$:
\begin{equation}
\frac{d\delta }{dt}=Kk^{2}\left[ -k^{2}\widetilde{\gamma }_{0}
+2k\frac{(\sigma_{xx}^{(0)}(C,C))^{2}}{M}- \frac{
d^{2}f_{el}^{(0)}}{dh^{2}}\right]_{h=C} \delta,
  \label{lin surface evolution}
\end{equation}
where $K=\frac{D_{s}\eta \Omega^{2}}{k_{B}T}$.
 Each term in the brackets in this equation has a simple physical
significance.
 The first term is a surface tension
term. Surface tension acts to reduce surface curvature, $\kappa$, and
so this term is negative, thereby reducing the perturbation amplitude,
and is linear in $\kappa \sim k^{2}$. The second term in this equation
is a mismatch stress term. Regions of high stress have large chemical
potential, and so atoms tend to detach from these regions and attach to
regions of small chemical potential. In a mismatch stressed solid,
valleys or cusps are regions of high stress, hence material moves from
the valleys to the hills of a perturbed surface increasing perturbation
amplitude. The contribution of this term is propotional to the density
of valleys, which is linear in $k$. The last term is a reference state
term. If $d^{2}f_{el}^{(0)}/dh^{2}>0$ it costs more energy to add a
monolayer to a flat film than to remove a monolayer, and hence it costs
energy to perturb a film. Thus, positive $d^{2}f_{el}^{(0)}/dh^{2}$
stabilizes a flat thin film, whereas negative
$d^{2}f_{el}^{(0)}/dh^{2}$ leads to an instability. Obviously this
reference state term is present even if the film is flat and hence is
independent of $k$.

 Equation (\ref{lin surface evolution}) implies that the flat film is
stable at all perturbation wavelengths as long as
\begin{equation}
\frac{\left[ \sigma_{xx}^{(0)}(C,C)\right] ^{4}}{M^{2}}\leq
\widetilde{\gamma }_{0}\left. \frac{d^{2}f_{el}^{(0)}}{dh^{2}}\right|
_{h=C}~, \label{crit}
\end{equation}
and the equality holds at the critical wetting layer thickness, where
perturbations of wavenumber $k=\left[ \sigma_{xx}^{(0)}(C,C)\right]
^{2}/(M\widetilde{\gamma }_{0})$ are marginal. $\widetilde{\gamma}_{0}$
is positive if $\theta =0$ is a surface seen in the equilibrium free
crystal \cite{herring}. As mentioned earlier $\widetilde{\gamma
}_{0}\rightarrow \infty $ at a perfect facet, and is large and positive
on a surface with a small miscut, as is the case for most of the
materials used in epitaxial films. Therefore, a linearly stable wetting
layer of finite thickness can exist only if
$d^{2}f_{el}^{(0)}/dh^{2}>0$. Note that for the wetting layer to have a
finite rather than an infinite thickness, $d^{2}f_{el}^{(0)}/dh^{2}$
must decrease to a value less than the l.h.s. of Eq. (\ref{crit}) as
the thickness of the film increases. $\sigma_{xx}^{(0)}(C,C)$ depends
linearly on the lattice mismatch $\varepsilon $, and hence the l.h.s.\
of (\ref{crit}) is proportional to $\varepsilon ^{4}$, while the
r.h.s.\ of (\ref{crit}) is proportional to $\varepsilon ^{2}$ due to
the dependence of $f_{el}^{(0)}$ on lattice mismatch. Therefore, if
$d^{2}f_{el}^{(0)}/dh^{2}>0$, the thickness of the wetting layer
increases with decreasing lattice mismatch and diverges in the limit
$\varepsilon \rightarrow 0$.

Note that the maximum thickness of a flat film which is stable to
infinitesimal perturbations is given by (\ref{crit}) when the equality
holds. A film slightly thicker is unstable to perturbations of
wavelength $\lambda=2\pi M \widetilde{\gamma _{0}}/\left[
\sigma_{xx}^{(0)}(C,C)\right] ^{2}$. For films which are nearly perfect
facets with small miscut angles these wavelengths are larger than the
typical sample size and so practically such perturbations will never
occur. However as will be explained in Section V, the film can be
nonlinearly unstable to smaller wavelength perturbations of a non-zero
amplitude at physically reasonable wavelengths. Hence the inequality in
(\ref{crit}) is only useful in predicting the stability of films with
large miscut angles or above the roughening temperature. At small
miscut angles the stability of the film to large perturbations will
predict its maximum thickness. This issue is discussed in more detail
in Section V.
\section{Calculation of the nonlinear elastic free energy of a flat film}

As can be seen from both Eq. (\ref{dfdh3}) and Eq. (\ref{crit}), the
dependence of the nonlinear elastic free energy of a flat film $%
f_{el}^{(0)}(h)$ on film thickness, $h$, is vital in order to
determine both wetting layer thickness and thin film evolution.
This free energy depends strongly on the mismatch stress $\sigma
_{ij}^{(0)}$, and its dependence on the $y$ coordinate. As a
result of the sharp interface between the substrate and the film,
we expect $\sigma _{ij}^{(0)}$ to behave as a step function of $y$
with small corrections due to elastic relaxation. If we ignore
these small corrections, the resulting free energy
$f_{el}^{(0)}(h)$  is proportional to film thickness, and its
second derivative vanishes. Hence according to Eq. (\ref{crit}),
the thickness of the critical wetting layer vanishes. The
correction due to elastic relaxation is therefore extremely
important. As discussed earlier, this correction vanishes within
linear elasticity theory. This led some investigators\cite{kukta}
to claim that the variation in free energy over the interface was
due to nonelastic effects, e.g. film-substrate material mixing
over the interface. However, we claim that this is not necessary,
since nonlinear elasticity can explain the corrections to the
step-function form of the free energy.

The nonlinear elastic free energy of the reference state,
$f_{el}^{(0)}(h)$, is calculated for a solid with a flat surface of
height $h$. Hence in order to calculate $df_{el}^{(0)}/dh,$ we
determine $f_{el}^{(0)}$ for flat solids of heights $h+\delta /2$ and
$h-\delta /2,$ and use the estimate
$df_{el}^{(0)}/dh=[f_{el}^{(0)}(h+\delta /2)-f_{el}^{(0)}(h-\delta
/2)]/\delta $. Ideally, first principles, substance specific
calculations
should be performed in order to evaluate $\sigma _{xx}^{(0)}(h,h)$ and $%
f_{el}^{(0)}(h)$, and we intend to carry out such calculations.
However, the qualitative general behaviour of $f_{el}^{(0)}(h)$
can be obtained from much simpler models. To demonstrate this
point we carried out the calculation for two-dimensional networks
of balls and springs of varying lattice-type and spring constants.
In these calculations $\delta $ is one monolayer, and
$f_{el}^{(0)}(h)$ is calculated at film thicknesses of integer
numbers of monolayers from no film up to 10 monolayers of film.
Values of $f_{el}^{(0)}(h)$ for film heights of fractional
monolayers are interprolated from the values calculated at integer
monolayer heights.

In the ball-and-spring model the balls are connected by springs
that obey Hooke's law. Note that this does not imply that the
stress-strain relationship of the ball-and-spring network is
linear (a discussion of this point can be found in Feynman's
Lectures \cite{feynman}). The natural spring length has a step
variation over the interface, and the balls are placed on a
lattice with the substrate lattice constant. Thus balls in the
substrate are connected by springs of their natural length,
whereas balls in the film are connected by springs that have
undergone a hydrostatic transformation strain and have length
larger than their natural length by a factor of $1+e$, where $e$
is the homogeneous strain in the film. The network was then
allowed to relax, with the film free to move in the $y$-direction,
and periodic boundary conditions being applied in the
$x$-direction to ensure that the system boundaries in this
direction were fixed to the natural substrate length.

\begin{figure}[h]
   \vspace{0.5cm}
   \epsfxsize=85mm
   \centerline{\epsffile{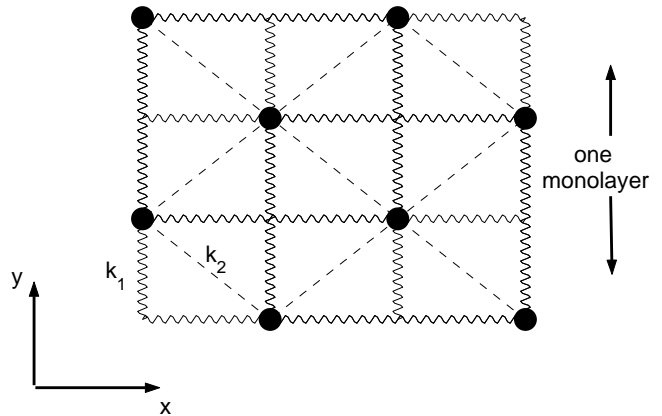}}
   \vspace{0.5cm}
\caption{Example of a fcc-like lattice. The circles represent balls.
The curvy lines are springs of spring constant $k_{1}$, and the dashed
lines represent springs of spring constant $k_{2}$.}
   \label{spring}
\end{figure}

We calculated the mismatch stress $\sigma_{xx}^{(0)}$ within the
relaxed film and at the film surface. We also calculated the dependence
of mismatch surface stress $\sigma_{xx}^{(0)}(h,h)$ and the nonlinear
elastic free energy $f_{el}^{(0)}(h)$ on varying film thickness, $h$.
We carried out these calculations for various two dimensional networks
of balls and springs with varying spring constants. An example of the
ball-and-spring model on a fcc-like lattice is shown in Fig.
(\ref{spring}).

\begin{figure}[h]
   \epsfxsize=85mm
   \centerline{\epsffile{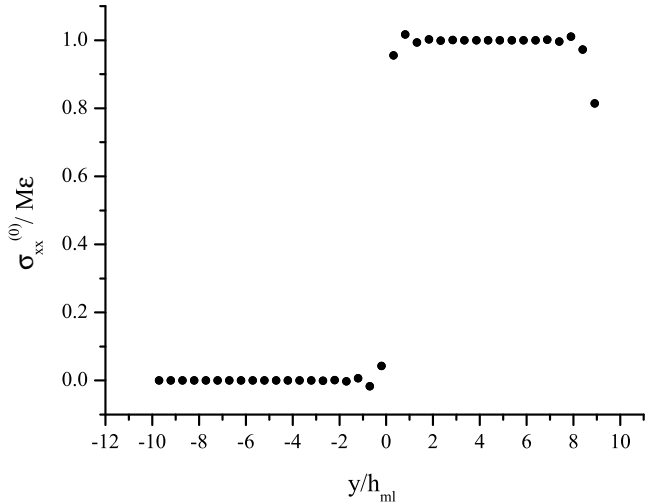}}
   \vspace{0.5cm}
\caption{Mismatch induced stress,
$\sigma_{xx}^{(0)}(h,y)/M\varepsilon$, in a typical fcc-like lattice.
The film is 8 monolayers thick and the substrate is 10 monolayers
thick. $h_{ml}$ is the thickness of one monolayer.}
   \label{stressinfilm}
\end{figure}

 Simulations showed that whilst
individual atoms were free to move in the $x$-direction, they
actually moved only in the $y$-direction. The relaxation in the
$y$-direction depended on the film thickness and on the depth of
the atom in the lattice but was independent of $x$. In general
balls at depth of more than 3 monolayers into the substrate
experienced no stress. The stress experienced by balls close to
the interface depended on the lattice type, spring constants and
ball position within the monolayer. A few monolayers into the
film, balls experienced the stress of an infinite thickness film,
$M\varepsilon$. At the film surface balls showed large relaxation.
Figure \ref{stressinfilm} shows an example of the mismatch stress,
$\sigma_{xx}^{(0)}(h,y)$ in a typical fcc-like lattice.

 Note that the springs in a simple square
lattice could relax completely in the $y$-direction. Therefore,
for such a lattice the relaxation is independent of spring depth
within the film or film thickness, and $f_{el}^{(0)}(h)$ varies
linearly with film thickness. Hence for a square lattice
$d^{2}f_{el}^{(0)}(h)/dh^{2}=0$. Only when diagonal bonds, such as
those in a fcc lattice were present did the springs show depth
dependent relaxation such as that described above. The inability
of the springs to completely relax due to the presence of diagonal
bonds was a necessary condition for $f_{el}^{(0)}(h)$ to vary
nonlinearly with film height. In such incompletely relaxed films
the nonlinear dependence of $f_{el}^{(0)}$ on $h$ arises from the
elastic relaxation at the surface and its coupling to the
relaxation at the film-substrate interface. A similar effect
should occur in real systems due to surface reconstruction, for
example.

\begin{figure}[h]
   \epsfxsize=75mm
   \centerline{\epsffile{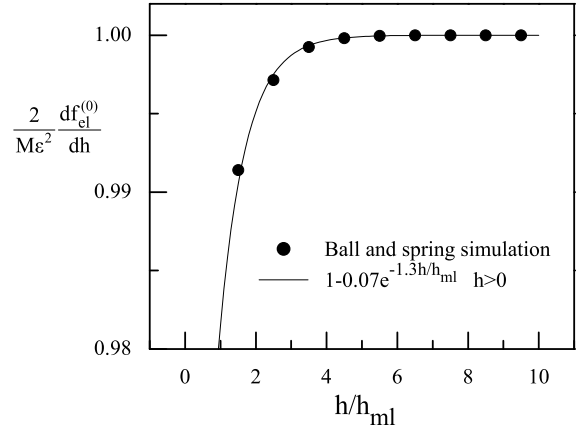}}
   \vspace{0.5cm}
\caption{Variation with film thickness of the elastic free energy of a
relaxed ball-and-spring system, $df_{el}^{(0)}/dh$, as a function of
film thickness $h$. The free energy is normalized to the infinite film
linear elastic energy density, $\frac{1}{2}M\varepsilon ^{2}$. $h_{ml}$
is the thickness of one monolayer.}
   \label{dfdhfig}
\end{figure}

A typical behavior of $df_{el}^{(0)}/dh$ is shown in Fig.
\ref{dfdhfig}, where it is seen that $f_{el}^{(0)}(h)$ indeed depends
on the thickness $h$. Moreover, the model predicts that
$d^{2}f_{el}^{(0)}/dh^{2}>0$ and decreases with increasing film
thickness, and therefore according to the inequality (\ref {crit}) and
the discussion following it, there should be a linearly stable wetting
layer, whose thickness is finite and increases with decreasing lattice
mismatch.

Whilst the detailed dependence of $df_{el}^{(0)}(h)/dh$ on film
thickness close to the substrate-film interface [$\leq 3\
\text{monolayers}$] varied between
different networks, it showed the same general behaviour. In all systems $%
d^{2}f_{el}^{(0)}(h)/dh^{2}$ showed exponential decay with a decay
length of
about a monolayer from the interface. The dimensionless quantity $\frac{2}{%
M\varepsilon ^{2}}\frac{df_{el}^{(0)}}{dh}$ was independent of lattice
mismatch sign or magnitude. $df_{el}^{(0)}/dh$ increased with film
thickness. As the film thickness increases $df_{el}^{(0)}/dh$
asymptotically approaches the elastic free energy density of an
infinite film, $M\varepsilon ^{2}/2 $, as expected.

We can explain the increase of $df_{el}^{(0)}/dh$ with film thickness
intuitively. $df_{el}^{(0)}/dh$ is the change in the free energy of a
film as a monolayer is added. When a monolayer is added to a thick film
the contribution to the free energy from the surface atoms and the
interface remains the same and the energy added is effectively that of
a monolayer in the 'bulk' of the film. Atoms in the bulk of a thick
film are more constrained than those in a thin film where the atoms are
relatively free to move and relax. When a monolayer is added to a thin
film the energy change is in between that of the constrained thick film
bulk atoms and the relaxed surface atoms. Hence more free energy is
needed to add a monolayer to a thick film and $df_{el}^{(0)}/dh$
increases with film thickness.

 For the calculations used later in this paper we used the
function
\begin{equation}
 df_{el}^{(0)}(h)/dh=\frac{M\varepsilon ^{2}}{2}[
1-0.05\exp (-h/h_{ml})] \;\;\;\;\;\text{for}\;\; \;\;h>0,
 \label{dfdhused}
\end{equation}
and $df_{el}^{(0)}(h)/dh=0$ for $h\leq 0$. $h_{ml}$ is the thickness of
one monolayer. The exponential decay form of $df_{el}^{(0)}(h)/dh$ fits
the results given by Tersoff \cite{tersoff}. In previous works
\cite{chiu,spencer,kukta} on the physics of the wetting layer it was
assumed that the reference state energy variation is a smooth function
of $h$, mainly in order to avoid non-analyticities at the interface. In
contrast, our reference state energy variation behaves as a step
function of the surface height with a small but important correction.

\begin{figure}[h]
   \centerline{
   \epsfxsize=40mm
   \epsffile{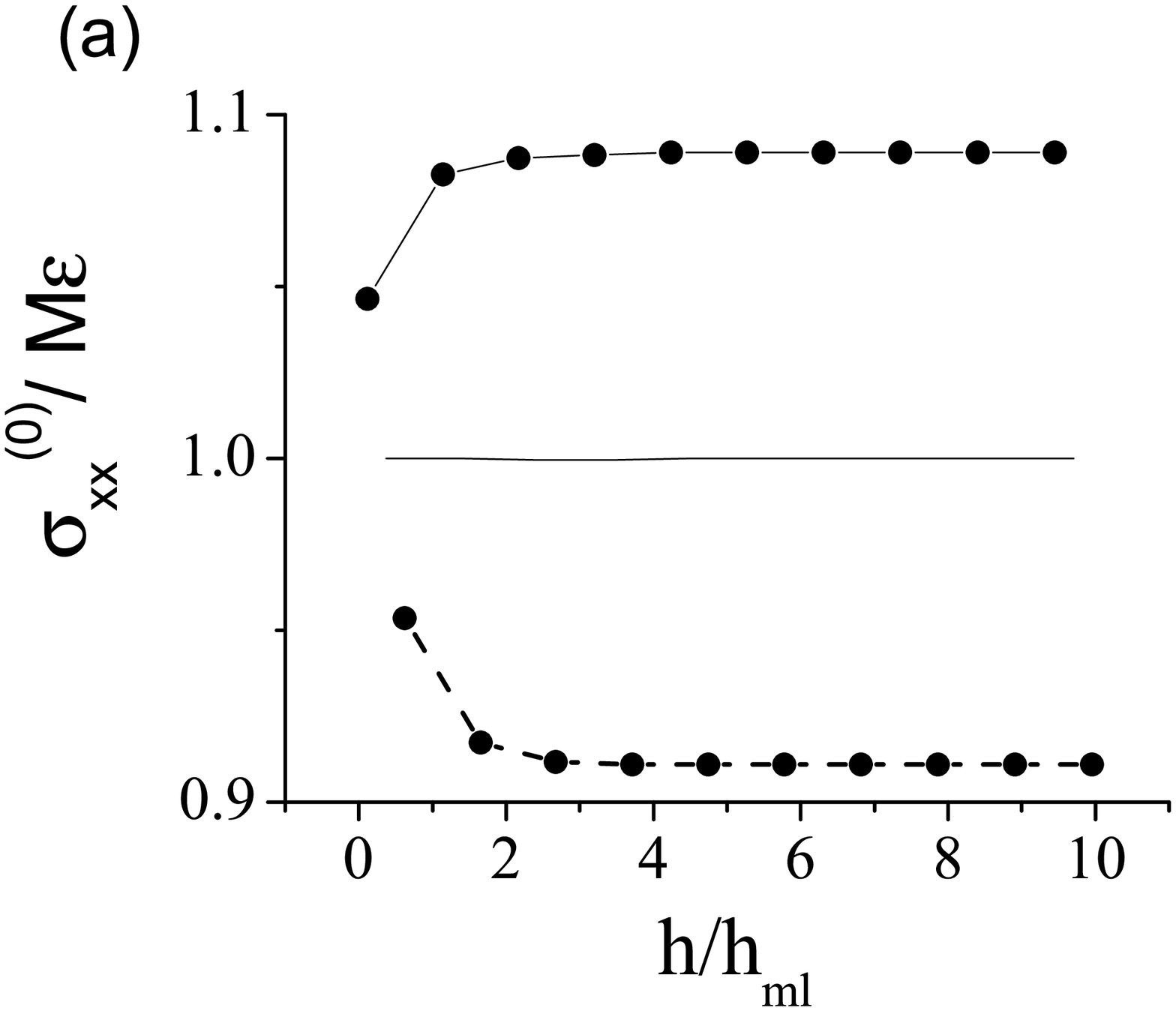}
   \hspace{-2mm}
   \epsfxsize=40mm
   \epsffile{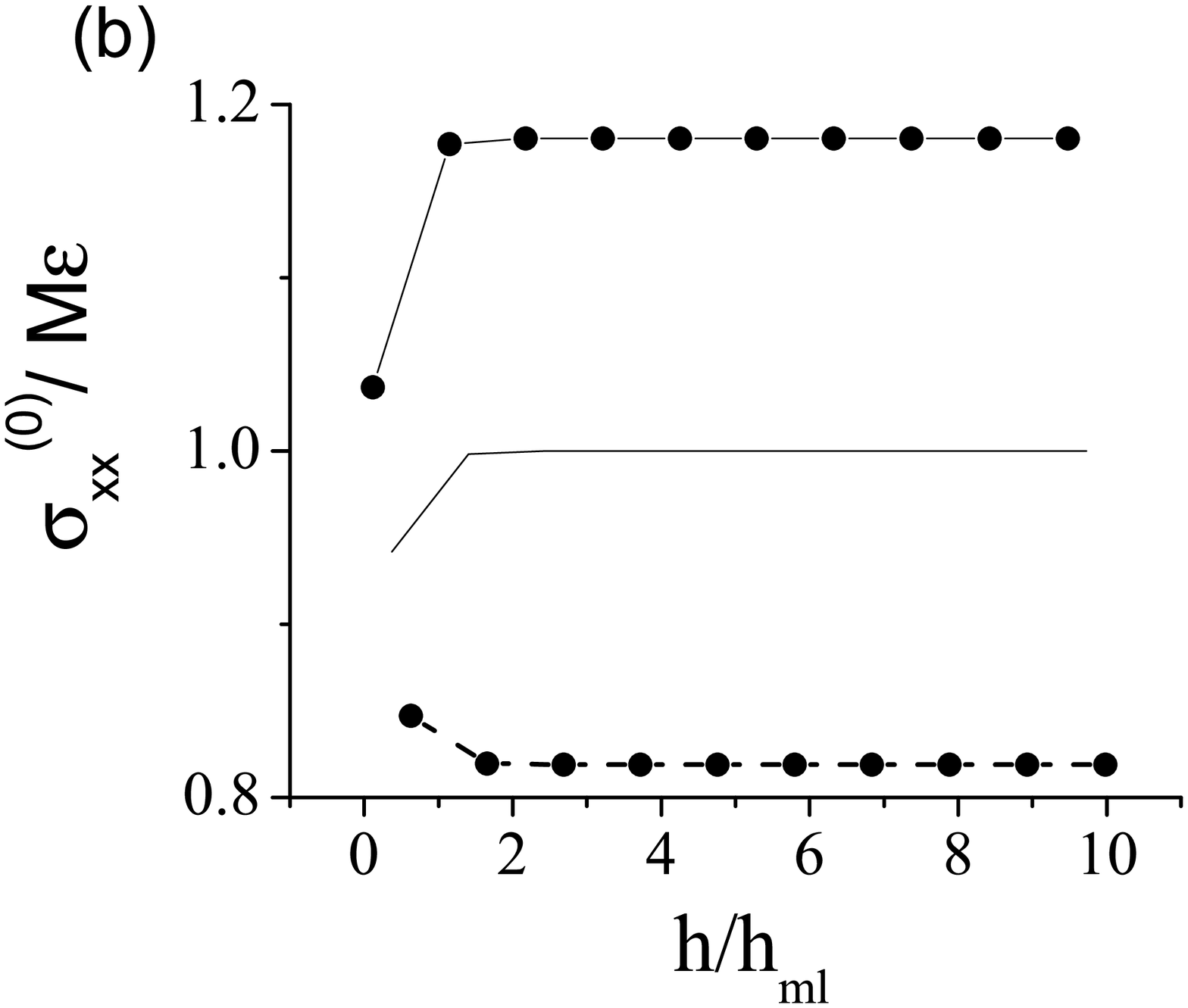}}
   \vspace{0.5cm}
\caption{Variation of mismatch induced stress at the film surface,
$\sigma_{xx}^{(0)}(h,h)/M\varepsilon$, for varying film thickness, $h$.
The lattice is fcc-like (see Fig. \ref{spring}). The dashed lines with
circles denote $\sigma_{xx}^{(0)}/M\varepsilon$ of atoms at the top of
a monolayer. The solid lines with circles denote
$\sigma_{xx}^{(0)}/M\varepsilon$ of atoms at the bottom of a monolayer.
The solid lines without circles represent the averages of
$\sigma_{xx}^{(0)}/M\varepsilon$ over a monolayer. The spring constants
which were used are (a) $k_{1}=1$ and $k_{2}=1$ and (b) $k_{1}=1$ and
$k_{2}=5$.}
   \label{surfacestress11}
\end{figure}

 The
deviations in the mismatch stress (averaged over the surface monolayer)
at the film surface $\sigma _{xx}^{(0)}(h,h)$ from a step function,
$\sigma _{xx}^{(0)}(h,h)=M\varepsilon $ when $h>0$, was shown to be
small ($<5\%$) but dependent on spring constants and lattice type (See
Fig. (\ref{surfacestress11})). As variations in $\sigma
_{xx}^{(0)}(h,h)$ only slightly alter the wetting layer thickness
predicted from Eq. (\ref{crit}), we decided to use the step function
form of mismatch stress.

Combining the behavior of $df_{el}^{(0)}/dh$ from Eq. (\ref{dfdhused})
with the inequality (\ref{crit}), we obtained an expression for the
linear stability wetting layer thickness, $h_c$:
\begin{equation}
h_{c}/h_{ml}=
\max\left\{1,ln\left[\widetilde{\gamma}_0/(40M\varepsilon^2
h_{ml})\right] \right\}~. \label{hc}
\end{equation}
Thus, the wetting layer thickness increases with decreasing lattice
mismatch, as observed in experiments.

\section{Numerical simulations of thin film growth}
As explained at the end of Section III, for systems with small mismatch
and/or small vicinality (miscut angle) one has to go beyond linear
stability in order to understand their evolution. To this end we
carried out numerical simulations of the evolution of the strained
film. The evolution equation (\ref{evol}) given by Mullins
\cite{mullins}, which is derived from the Nernst-Einstein relation
(\ref{nernst}), includes derivatives of the chemical potential, $\mu$,
along the surface. This chemical potential is defined as the change in
free energy when an atom is added to the surface. Continuum theory
assumes that the free energy change when the surface is changed by an
infinitesimal amount is proportional to this chemical potential
$(\mu=\Omega\frac{\delta F}{\delta h})$. However, when we solved the
evolution equation by directly calculating the chemical potential from
Eq. (\ref{dfdh3}) at points along the film surface we experienced
numerical instabilities.

We have come up with the following solution to this problem. The
Nernst-Einstein equation can be derived by considering material of
atomic volume moving along the solid surface. When material jumps
between neighboring atomic sites, it must cross a free energy barrier
of $\Delta F_{\pm}=E_{d}+(F_{\pm}-F_{0})/2$, where $E_{d}$ is the
potential barrier for diffusion, $F_{\pm}$ is the free energy of the
film after material has been moved, and $F_{0}$ is the free energy of
the film before material is moved. The positive and negative signs
stand for forward and backward jumps, respectively. This leads to the
following equation for material velocity along the surface
\begin{eqnarray}
v&=&\omega a
e^{-E_{d}/k_{B}T}(e^{-(F_{+}-F_{0})/2k_{B}T}-e^{-(F_{-}-F_{0})/2k_{B}T})
\nonumber \\
&=&\frac{D_{s}}{a}(e^{-(F_{+}-F_{0})/2k_{B}T}-e^{-(F_{-}-F_{0})/2k_{B}T}),
\label{nernstnonlin}
\end{eqnarray}
 where $\omega$ is the attempt rate and $a$ is the jump length.
  When $\Delta F_{\pm}$ is small, this equation gives
  the Nernst-Einstein relation
(\ref{nernst}).

We solved Eq. (\ref{nernstnonlin}) using the following numerical
scheme. For every two adjacent points on the surface, the surface
height of the left point was changed by $\pm \delta h$ and of the right
point by $\mp \delta h$ so as to give a transfer of material of atomic
volume backwards and forwards along the surface respectively. The
change in surface free energy, $\int dx\ \gamma \sqrt{1+(\partial
h/\partial x)^{2}}$, was calculated for this material transfer. The
change in the elastic free energy was calculated at each point using
the integral of Eq. (\ref{dfdh3}),
$\delta F=\pm\left[\frac{\ df_{el}^{(0)}}{%
dh}+\left. \left( \frac{1}{2}S_{ijkl}\sigma _{ij}\sigma _{kl}-\frac{1}{2}%
S_{ijkl}\sigma _{ij}^{(0)}\sigma _{kl}^{(0)}\right) \right|
_{y=h(x)}\right]\delta h$. The linear elastic energy was calculated by
solving the biharmonic equation (\ref{biharmonic}) with the boundary
conditions (\ref{bound}). This was done by solving a boundary integral
equation in terms of the complex Goursat function, the details of which
can be found in the paper of Spencer and Meiron \cite{spencermeiron}.

\subsection{The stability of thin films}

\begin{figure}[h]
   \epsfxsize=75mm
   \centerline{\epsffile{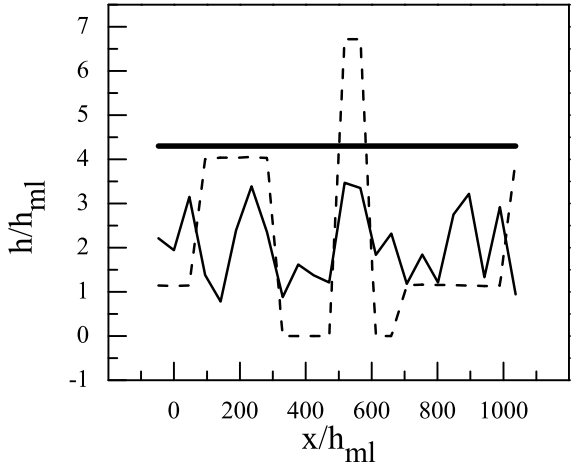}}
   \vspace{0.5cm}
   \caption{ Evolution of a randomly perturbed film, in which
perturbations were larger than the critical perturbation amplitude.
Lattice mismatch in this film is 4\%. The initial film surface is shown
as a thin solid line. The dashed line shows the film surface at a later
time. The linear wetting layer thickness is shown as a thick solid
line. }
   \label{filmevolfig}
\end{figure}

According to Eq.\ (\ref{hc}), anisotropic surface tension greatly
enlarges the linearly stable wetting layer thickness. Does this
conclusion survive beyond linear stability analysis? When a
linearly stable flat film is perturbed strongly so that the
surface orientation in some regions is far from the $\theta =0$
direction, the local surface stiffness in these regions is much
smaller than the $\theta =0$ stiffness. This tends to destabilize
the linearly stable film. Indeed, we carried out numerical
simulations (using the procedure described above) that showed that
films thinner than the linear wetting layer were unstable to
perturbations greater than a certain critical amplitude (see Fig.
\ref{filmevolfig}). Hence films thinner than the linear wetting
layer thickness are {\em metastable}. When large perturbations
were applied, faceted islands developed in the film, which
underwent ripening at later stages of the evolution.

\begin{figure}[h]
   \epsfxsize=80mm
   \centerline{\epsffile{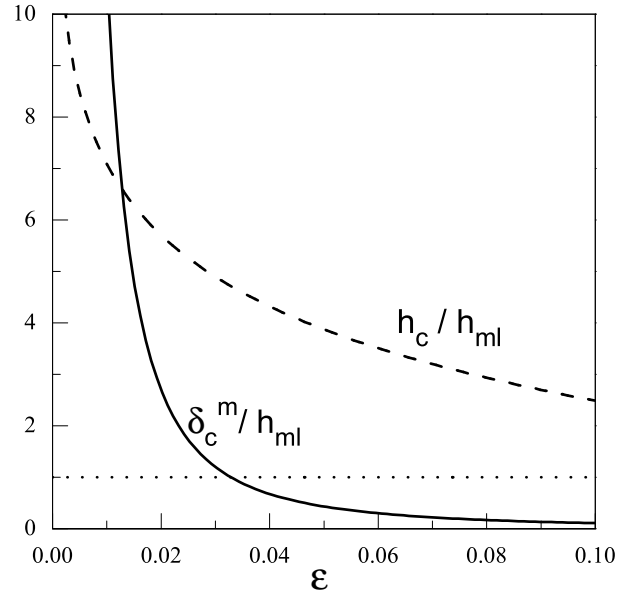}}
   \vspace{0.5cm}
\caption{Variation of the minimal critical perturbation amplitude,
$\delta_{c}^{m}$, and linear wetting layer thickness, $h_{c}$, with
lattice mismatch, $\varepsilon$. The minimal critical perturbation
amplitude, $\delta_{c}^{m}/h_{ml}$ is represented by the solid line.
The linear wetting layer thickness, $h_{c}/h_{ml}$, is represented by
the dashed line. The dotted line shows the size of one monolayer,
$h_{ml}$, for comparison.}
   \label{critpert}
\end{figure}

We carried out simulations for films perturbed by random perturbations
and by perturbations of a single wavelength. The critical perturbation
amplitude, $\delta_{c}$, depends on the wavelength of the perturbation,
$\lambda$, taking its minimal value
\begin{equation}
\delta_{c}^{m}=\min_\lambda \delta_{c}(\lambda)
\end{equation}
 at
$\lambda/l_{o}\sim10-50$, where $l_{o}=2\gamma_{o}/M\varepsilon^{2}$.
$\delta_{c}^{m}$ in monolayers is plotted as a function of lattice
mismatch in Fig. \ref{critpert}. The linear wetting layer thickness for
$G=500$, $M=1.5\times 10^{11} N/m^2$ and $h_{ml}=5{\mbox \AA}$ is also
shown for comparison.

$\delta_{c}^{m}$ was found to be proportional to $\varepsilon
^{-2}$. The $\varepsilon ^{-2}$ dependence is expected for an
infinite film as in this case the evolution equations (Eq.
(\ref{evol}) together with Eq. (\ref{dfdh3})) can be
 made spatially dimensionless by scaling all lengths by $l_{0}$. Hence all
   perturbations of size $\delta/l_{0}$ and with the same
   dimensionless wavenumber
  $kl_{0}$ will evolve
   identically.

$\delta_{c}^{m}$ was largely independent of cusp smoothness $G$,
unlike the linear wetting layer thickness which depended strongly
on $G$. This suggests that unlike the linear wetting layer
thickness the critical perturbation amplitude can be used in
predicting the outcome of experimental thin film growth. The mean
square amplitude of the random perturbation needed to destabilize
thin films was also shown to be largely independent of $G$ and was
proportional to $\varepsilon ^{-2}$.

When the lattice mismatch is small, $\delta_{c}^{m}$ is much larger
than the linear wetting layer thickness (see Fig. \ref{critpert}).
Hence, flat films thinner than the linear critical thickness are stable
at small lattice mismatch. As the linear critical thickness at small
lattice mismatch is very large, we expect the film to first become
unstable to misfit dislocations. This in indeed seen in experiments
\cite{tromp,perovic}.

 At intermediate lattice mismatch, $\delta_{c}^{m}$ is of the order of
a few monolayers.
  Hence we expect
such a film to become unstable as perturbations of this amplitude
are physically likely. In this regime films should develop growing
perturbations at wavelengths given by $\lambda/l_{o}\sim10-50$.
This corresponds to wavelengths of a few hundred nanometers. This
typical wavelength decreases as lattice mismatch increases,
agreeing with experiment \cite{floro1997,tromp}. As $\varepsilon$
increases, $\delta_{c}^{m}$ decreases in this regime from about 10
monolayers to approximately one monolayer, we expect the thickness
of the film needed to support such perturbations to
correspondingly decrease. Such a trend is seen in experiments
\cite{floro1997,floro1999,tromp,perovic}. In order to compare
quantitative wetting layer thickness and its lattice mismatch
dependence with experiments we will have to carry out first
principles, substance-specific calculations to evaluate
$f_{el}^{(0)}(h)$. However, the general qualitative trends
predicted here agree with experimental observations.

Looking at Fig. \ref{critpert} we see that at intermediate lattice
mismatch the critical perturbation amplitude, $\delta_{c}^{m}$, and the
linear wetting layer thickness are of the same order of magnitude
(several monolayers). This could mean that we have to also consider the
linear wetting layer thickness with its strong dependence on the
surface miscut angle when deciding whether or not a thin film will be
unstable. However, the infinitesimal perturbations needed to perturb
the linear wetting layer occur at
 wavelength,
$\lambda=\frac{2\pi}{M\varepsilon^{2}}\widetilde{\gamma}_{0}$, whereas
the typical wavelength at which critical amplitude perturbations first
appear in the film is,
$\lambda\sim10\frac{2}{M\varepsilon^{2}}\gamma_{0}$. Using expression
(\ref{surf ten}), the ratio between these two wavelengths is
approximately equal to $G$. For physical values of $G$ the linear
wetting layer perturbations will have wavelengths of the order of
$100\mu m$ which is larger than the typical sample size, whereas the
wavelength which corresponds to the minimal critical
 perturbation is much smaller ($\sim100nm$).
 Therefore physical thin films should first become unstable
when the film thickness is large enough to support perturbations larger
than the critical perturbation amplitude.

 For very large mismatch, a perturbation smaller than a monolayer is
sufficient in order to destabilize the linearly stable wetting layer.
Therefore, in practice, the wetting layer is a single monolayer in this
case.

\section{Early evolution of thin films with material
deposition}

We carried out our calculations with two different types of material
deposition: The first type is deposition at a steady rate in
 the vertical $y$-direction, corresponding to any directed deposition
(e.g, molecular
  beam epitaxy). The evolution equation (\ref{evol})
  then becomes
 \begin{equation}
\frac{\partial h}{\partial t}=\frac{D_{s}\eta\Omega }{k_{B}T}
\frac{\partial }{\partial x}\frac{
\partial \mu }{\partial s}+V_{D}, \label{depevol}
\end{equation}
where $V_{D}$ is the material deposition rate.

The second type is deposition constant in the direction perpendicular
to the film surface, corresponding to liquid phase epitaxy, for
example. Early growth with this method of deposition has been studied
by Chiu and Gao \cite{chiu}. In this case the evolution equation
becomes
\begin{equation}
\frac{\partial h}{\partial t}=\frac{D_{s}\eta\Omega }{k_{B}T}
\frac{\partial}{\partial x}\frac{%
\partial \mu }{\partial s}+\frac{V_{D}}{n_{y}},\label{depevol1}
\end{equation}
where $n_{y}$ is the $y$-component of the normal vector to the surface.

 We performed linear stability analysis in order to obtain the
analytical early evolution equation of a perturbed thin film. This
analysis is valid for both types of material deposition. Under steady
deposition, Eq. (\ref{lin surface evolution}) becomes
\begin{equation}
\frac{d\delta }{dt}=K\left[ -k^{4}\widetilde{\gamma
}_{0}+2k^{3}M\varepsilon^{2}-k^{2} \frac{
d^{2}f_{el}^{(0)}(C+V_{D}t)}{dh^{2}} \right] \delta ~,
\end{equation}
where we have assumed the reference state mismatch stress is given by a
step function, $\sigma _{xx}^{(0)}(h,h)=M\varepsilon $ when $h>0$, and
$\sigma _{xx}^{(0)}(h,h)=0 $ when $h<0$.
 Using the general form (obtained from the ball-and-spring model)
  of $d^{2}f_{el}^{(0)}/dh^{2}\simeq
\chi \frac{M\varepsilon^{2}}{2h_{ml}}\exp(-h/h_{ml})$, where $\chi$ is
a constant, gives the following solution for perturbation growth:

\begin{eqnarray}
&\delta&(t)=
 \delta_{0}\exp\left\{K\left[(-k^{4}\widetilde{\gamma
}_{0}+2k^{3}M\varepsilon^{2})t + \right.\right.\nonumber \\
&&\left.\left.k^{2}\chi (M\varepsilon^{2}/2V_{D})
exp(-C/h_{ml})(exp(-V_{D}t/h_{ml})-1)\right]\right\}. \label{depgrowth}
\end{eqnarray}

Note that in linear stability analysis a perturbation in an infinite
film decays when $k>2M\varepsilon^{2}/\widetilde{\gamma }_{0}$ and
grows exponentially when $k<2M\varepsilon^{2}/\widetilde{\gamma }_{0}$.

\begin{figure}[h]
   \centerline{
   \hspace{-2mm}
   \epsfxsize=40mm
   \epsffile{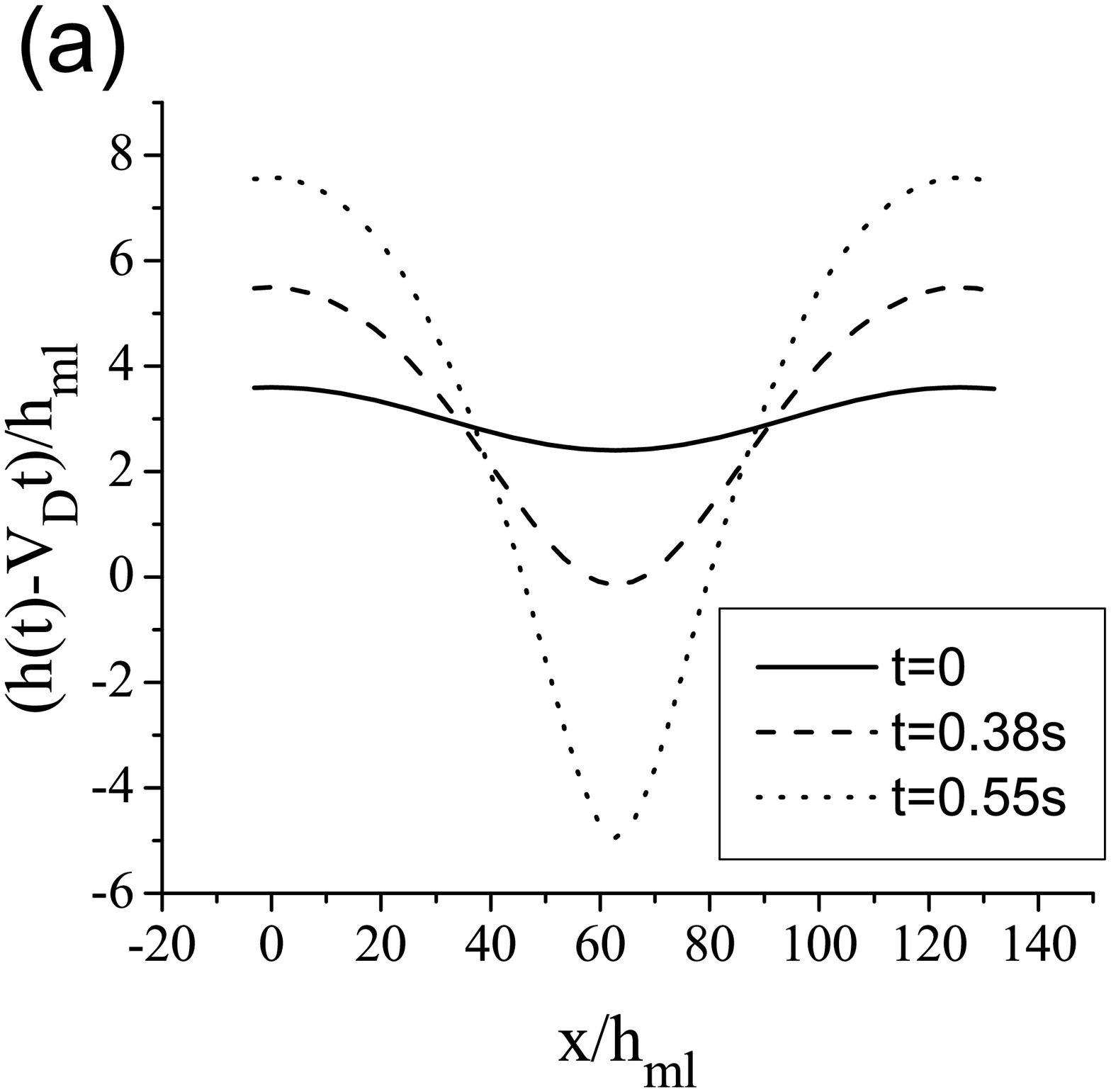}
   \hspace{-0mm}
   \epsfxsize=40mm
   \epsffile{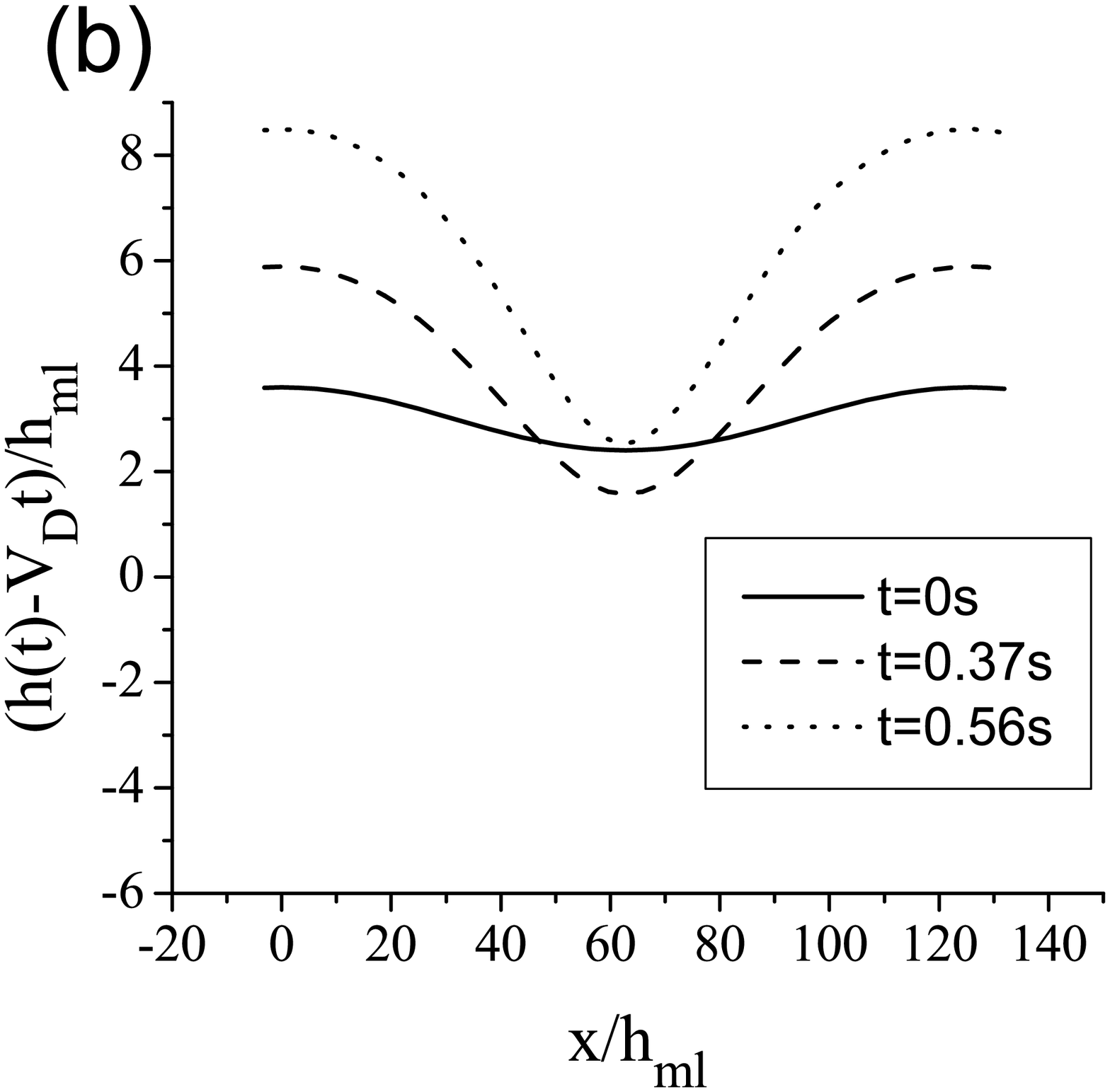}}
   \vspace{0.5cm}
\caption{Film evolution at very high deposition rates
($V_{D}=1200nm/s$) when surface tension is isotropic. $k=0.75\times
2M\varepsilon^{2}/\widetilde{\gamma }_{0}$. (a): Deposition is constant
in the vertical $y$-direction. (b): Deposition is constant
perpendicular to the film surface.}
    \label{filmevol}
\end{figure}

 For isotropic surface tension, numerical computations showed
that when $k^{*}<k<2M\varepsilon^{2}/\widetilde{\gamma }_{0}$, with
$k^{*}\approx 0.875\times 2M\varepsilon^{2}/\widetilde{\gamma }_{0}$,
both methods of deposition lead to cusp formation in the surface
valleys. The cusps initially evolve according to the linear evolution
equation (\ref{depgrowth}) and then slow and reach a steady state
morphology. Spencer and Meiron \cite{spencermeiron} observed such
steady states in infinitely thick stressed films. However when
$k<k^{*}$, surface evolution depends on the method of material
deposition. When deposition is constant in the vertical $y$-direction
increasingly sharp cusps form in the surface valleys (see Fig.\
\ref{filmevol}a), which continue to grow exponentially. In contrast
when deposition is constant perpendicular to the surface at very high
deposition rates cusp formation is slowed (see Fig.\ \ref{filmevol}b)
and the surface shows signs of reaching a steady-state morphology as
for $k>k^*$.

\begin{figure}[h]
   \epsfxsize=85mm
   \centerline{\epsffile{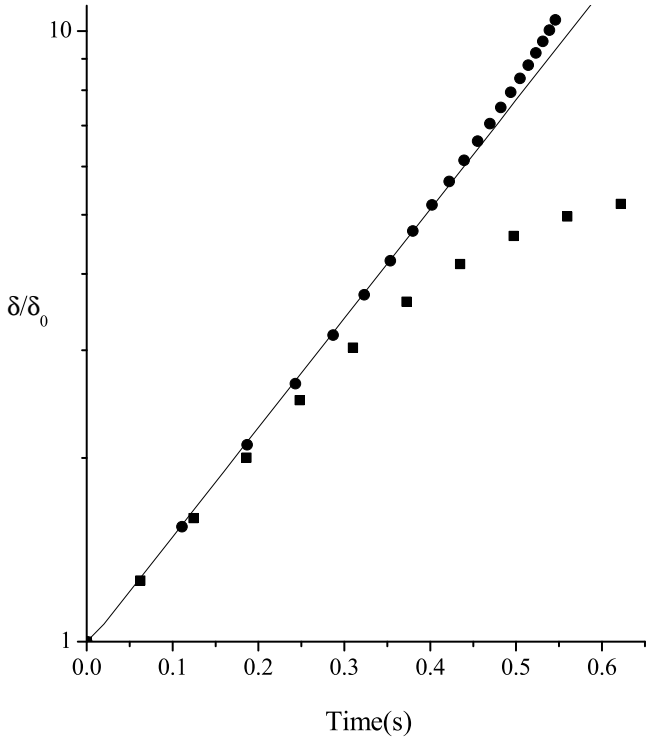}}
   \vspace{0.5cm}
\caption{Film evolution at very high deposition rates
($V_{D}=1200nm/s$) when surface tension is isotropic. The line
represents the linear evolution equation, the circles represent the
results from the numerical simulation when deposition is constant in
the $y$-direction and the squares represent the results from the
numerical simulation when deposition is constant perpendicular to the
surface. $k=0.75\times 2M\varepsilon^{2}/\widetilde{\gamma }_{0}$. The
parameters used here are: $D_{s}=3.599\times10^{-13}m^{2}/s$,
$\Omega=1.38\times10^{-29}m^{3}$, $\eta=1.74\times10^{19}m^{-2}$,
$T=700K$, $k=10^{8}m^{-1}$, $\gamma_{0}=\widetilde{\gamma }_{0}=1$,
$\varepsilon=2\%$, $M=1.67\times10^{11}$, $\chi=1$, $C=0.75$ monolayers
and $h_{ml}=2nm$. Note the deviation from the linear stability analysis
results when deposition is perpendicular to the surface.}
    \label{isogrowth2}
\end{figure}

This can be seen in Fig. \ref{isogrowth2}. The plot (shown as squares
in the figure) starts as a graph of constant positive slope
representing an exponentially growing perturbation as predicted by
linear analysis. However this growth slows and the graph approaches the
flat line representative of a steady-state morphology. Note that in
comparison when deposition is constant in the vertical $y$-direction
(shown as circles in Fig. \ref{isogrowth2}), the film evolves according
to the linear evolution equation. Under deposition perpendicular to the
film surface, when a cusp begins to form, material is deposited more
rapidly on the steep cusp slope, hence slowing cusp formation. When
deposition is constant vertically it only effects surface evolution
indirectly by raising the average surface height. Though deposition
rates of this magnitude are not generally used in experiment it is
nevertheless physically interesting to observe the difference in
surface evolution between the two growth methods.

\begin{figure}[h]
   \epsfxsize=85mm
   \centerline{\epsffile{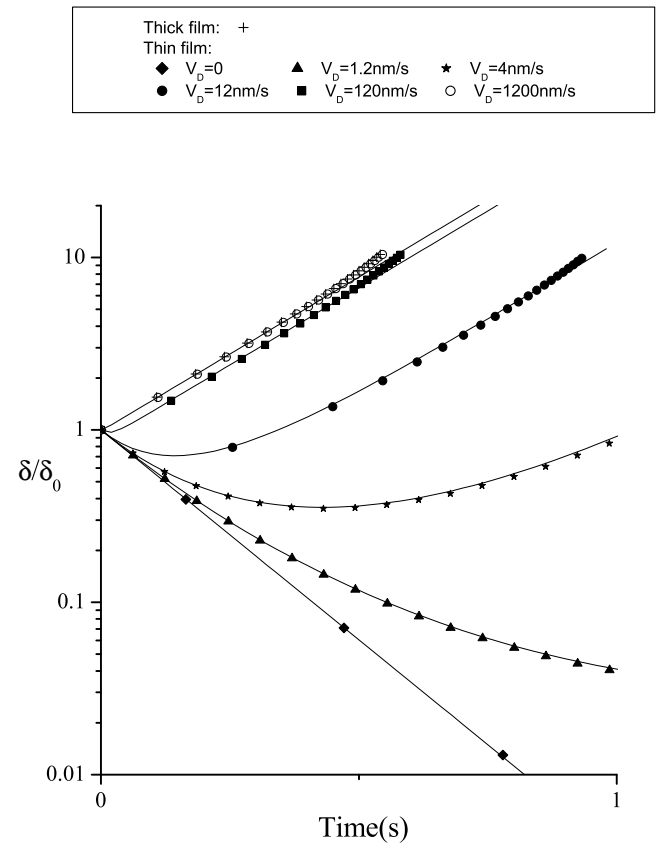}}
   \vspace{0.5cm}
\caption{Film growth when surface tension is isotropic and deposition
is constant in the $y$-direction. The symbols represent the numerical
simulation results and the lines the linear evolution equation values.
$k=0.75\times 2M\varepsilon^{2}/\widetilde{\gamma }_{0}$. The
parameters used are the same as those in Fig. \ref{isogrowth2}.}
    \label{isogrowth1}
\end{figure}

When the deposition is constant in the vertical $y$-direction, the film
evolves according to the linear evolution equation, even after the
surface is no longer a sine function and cusp formation occurs. This
can be seen in Fig. \ref{isogrowth1} which compares results from the
numerical simulation with the results predicted by the linear evolution
equation (\ref{depgrowth}). Figure \ref{filmevol} shows a very clear
cusp formation in the surface morphology, whilst for the same time Fig.
\ref{isogrowth1} shows the sharp cusp growing only slightly faster than
predicted by linear analysis. This slight deviation is expected as the
stress in a cusp valley is larger than in a sine valley hence
accelerating perturbation growth.

 When the surface tension is anisotropic the surface
evolution is very different from that predicted by the linear
analysis. As can be seen in Fig. \ref{isogrowth1}, a perturbation
in an isotropic film decays until the film surface reaches a
height at which the film is linearly unstable to perturbations at
that wavelength . No matter how large the deposition rate, at any
given time a perturbation in a thin film is always smaller than a
perturbation of the same initial size in an infinite film due to
the finite time spent in the linear wetting layer.

 When
surface tension is anisotropic, initially the perturbation
amplitude decreases as the surface facets. The rate and amplitude
of the decrease is independent of either $C$ or $V_{D}$. The film
then grows or decays depending on whether the perturbation
amplitude $\delta(t)$, is larger or smaller than the critical
perturbation amplitude $\delta_{c}(k,h)$, at $h=V_{D}t+C$. When
$\delta_{0}>\delta_{c}(k,h=C)$ the perturbation grows immediately
after faceting. When $\delta_{0}<\delta_{c}(k,h=C)$ the
perturbation initially decays though it will start to grow again
if the deposition brings the film surface to a height at which
$\delta(t)>\delta_{c}(k,h=V_{D}t+C)$.

We now turn to characterize the evolution of the film in terms of all
the relevant physical variables. There are five independent variables
that can effect the evolution: $V_{D},C,k,t$ and
 $\varepsilon$. In addition, there are four relevant constant
parameters: $h_{ml},K,\widetilde{\gamma }_{0}$ and $M$. We can replace
the five independent variables by the dimensionless variables:
$V_{D}t/h_{ml},\;C/h_{ml},\;K\widetilde{\gamma
}_{0}k^{4}t,\;KM\varepsilon^{2}k^{3}t,\;KM\varepsilon^{2}k^{2}t/h_{ml}$.
The idea is that the description of the evolution of a film with
isotropic surface tension becomes simpler in terms of these variables.
This becomes clear when we look at how a perturbation in a thin film
grows in relation to the growth of a perturbation of the same initial
size in an infinite film. Quantitatively, this is described by the
relative perturbation height,
$\delta_{R}=\delta(t,C)/\delta(t,C=\infty)$. As can be seen from Eq.
(\ref{depgrowth}), $\delta_{R}$ depends only on three of the five
independent dimensionless variables:
\begin{equation}
 \delta_{R}=e^{Kk^{2}\chi
\frac{M\varepsilon^{2}}{2V_{D}}exp(-C/h_{ml})(exp(-V_{D}t/h_{ml})-1)},
\label{reldepgrowth}
\end{equation}
 This can be summarized in the scaling law:
\begin{equation}
\delta_{R}(V_{D},C,k,t,\varepsilon)
=\delta_{R}(V_{D}t/h_{ml},C/h_{ml},KM\varepsilon^{2}k^{2}t/h_{ml}),
\label{scaling}
\end{equation}
which implies that $\delta_{R}$ depends only on three of the five
scaling variables. The manifestation of this scaling behaviour is data
collapse. For example, when $V_{D}=0$ and $C$ is fixed, all the curves
of $\delta _{R}(t)$ for different values of $k$ and $\varepsilon$ fall
onto a single curve if plotted as a function of $k^{2}\varepsilon^{2}t$
rather than $t$.

 Does this scaling law survive beyond linear
analysis? We looked for scaling when deposition was constant in the
vertical $y$-direction for both isotropic and anisotropic surface
tension. As mentioned earlier when surface tension is isotropic the
film continued to evolve according to the linear evolution equation
(\ref{depgrowth}) long after it left the linear regime and hence the
scaling relation (\ref{scaling}) also held.

\begin{figure}[h]
   \epsfxsize=80mm
   \centerline{\epsffile{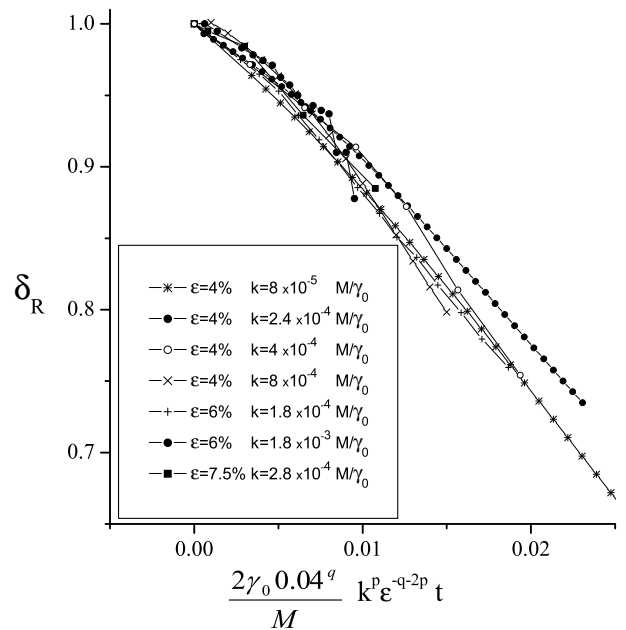}}
   \vspace{0.3cm}
\caption{Scaling of relative perturbation height for different $k$ ($k=
8\times10^{-5}\frac{M}{\gamma_{0}} \rightarrow
2\times10^{-3}\frac{M}{\gamma_{0}}$) and $\varepsilon$
($\varepsilon=4\%\rightarrow7.5\%$) with zero deposition.
 $p=2.37$ and $q=-6.5$. Note that the variable used to plot this graph
 $\frac{2\gamma_{0}0.04^{q}}{M}k^{p}\varepsilon^{-2p-q}t$ is
 proportional to the scaling variable $K \widetilde {\gamma
}_{0}^{p+q/2+1}M^{-p-q/2}\varepsilon^{-2p-q}k^{p}t/h_{ml}^{q/2+4}$.}
   \label{scaling1}
\end{figure}

\begin{figure}[h]
   \epsfxsize=85mm
   \centerline{\epsffile{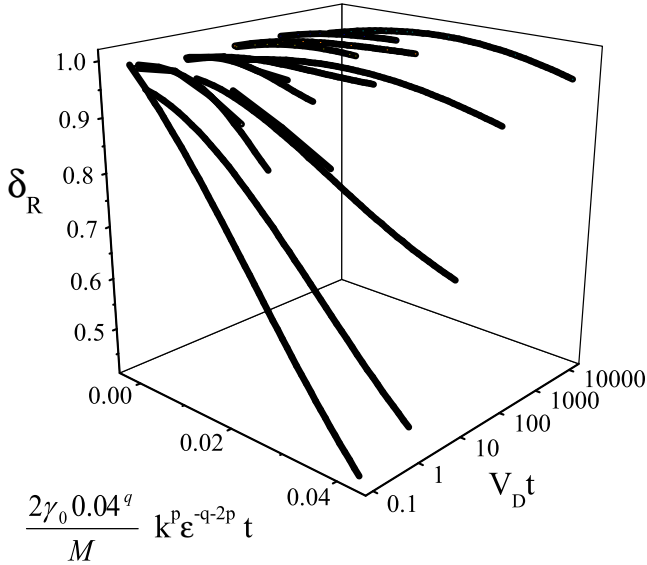}}
   \vspace{0.5cm}
\caption{Scaling of relative perturbation height for different $k$
($k=8\times10^{-5}\frac{M}{\gamma_{0}} \rightarrow
2\times10^{-3}\frac{M}{\gamma_{0}}$) and $\varepsilon$
($\varepsilon=4\%\rightarrow7.5\%$) and $V_{D}$
($V_{D}=0,0.12,1.2,12,120,1200,12000,120000\AA/s$).
 $p=2.37$ and $q=-6.5$. Note that the variables used to plot this graph
 $\frac{2\gamma_{0}0.04^{q}}{M}k^{p}\varepsilon^{-2p-q}t$ and $V_{D}t$
 are
 proportional to the scaling variables $K \widetilde {\gamma
}_{0}^{p+q/2+1}M^{-p-q/2}\varepsilon^{-2p-q}k^{p}t/h_{ml}^{q/2+4}$ and
$V_{D}t/h_{ml}$. }
     \label{scaling3}
\end{figure}

Growth under anisotropic surface tension is very different from that
given by the linear evolution equation (\ref{depgrowth}), and hence the
scaling relation (\ref{scaling}) does not hold. Does this mean that the
physics of anisotropic surfaces is more complicated and depends on all
five independent variables? It turns out the answer to this question is
\textit{no} to a good approximation. To see this we define five
generalized dimensionless variables: $V_{D}t/h_{ml}$, $C/h_{ml}$, $K
\widetilde {\gamma }_{0}k^{4}t,$ $KM\varepsilon^{2}k^{3}t,$ and $K
\widetilde {\gamma
}_{0}^{p+q/2+1}M^{-p-q/2}\varepsilon^{-2p-q}k^{p}t/h_{ml}^{q/2+4}.$
 When $p=2$ and $q=-6$ we regain
the dimensionless variables governing the linear evolution equation. We
found numerically that in the case of anisotropic surface tension,
$\delta_{R}$ approximately obeys the scaling law: :
\begin{eqnarray}
\delta_{R}(k,V_{D},&\varepsilon&,t,C)
=\delta_{R}(V_{D}t/h_{ml},\,C/h_{ml},
\nonumber \\
&K& \widetilde {\gamma
}_{0}^{p+q/2+1}M^{-p-q/2}\varepsilon^{-2p-q}k^{p}t/h_{ml}^{q/2+4})
 \label{genscal}
\end{eqnarray}
with $p\sim2.37$ and $q\sim-6.5$. Again, $\delta_{R}$ depends only on
three of the five scaling variables, which implies data collapse.
 This relation was very robust. We verified it for
different G, variation of $k$ of an order of magnitude, variation
of $\varepsilon$ by 100\% and deposition rates of between 0 and
120000\AA/s. Figures \ref{scaling1} and \ref{scaling3} show this
scaling in the form of data collapse when $C=2ML$. Data collapse
when $\delta_{R}$ is plotted against a variable proportional to
the third scaling variable in Eq. (\ref{genscal}) can be seen in
Fig. \ref{scaling1}. Here there is no deposition, $k$ varies by
over an order of magnitude and $\varepsilon$ by 100\%. As can be
seen the data collapse is not exact but holds to a good
approximation. Figure \ref{scaling3} shows data collapse when
$\delta_{R}$ is plotted against variables proportional to the
first and third scaling variables in Eq. (\ref{genscal}).
Deposition rates vary by six orders of magnitude, $k$ varies by
over an order of magnitude and $\varepsilon$ by 100\%. Data
collapse is shown by all curves falling onto a single surface.

The scaling relationship (\ref{genscal}), however, only held when
the initial perturbation $\delta_{0}$ was larger than the critical
perturbation at that wavenumber, $k$, and initial film height,
$C$; i.e for $\delta_{0}>\delta_{c}(k,C)$. This is probably
because when $\delta_{0}>\delta_{c}(k,C)$ perturbations of a thin
film and of an infinite film evolve similarly. Both perturbations
initially decay whilst faceting and then continue to grow. On the
other hand, when $\delta_{c}(k,\infty)<\delta_{0}<\delta_{c}(k,C)$
a perturbation of a thin film decays whereas an infinite film
perturbation facets and grows. In this regime scaling laws were
not found. When $\delta_{0}<\delta_{c}(k,\infty)$ the perturbation
decays in both the thin and infinite film. When $\delta_{0}$ is
small enough the scaling laws (\ref{scaling}) derived from the
linear evolution equation are regained.

We would like to thank J.Tersoff, B.J.Spencer and V.I.Marchenko for
interesting discussions relating to matter contained in this paper.

\end{document}